\documentclass[12pt]{article}

\usepackage[utf8]{inputenc}
\usepackage[T1]{fontenc}
\usepackage[english]{babel}
\usepackage{lmodern}
\usepackage{color}
\usepackage[round,authoryear]{natbib}
\usepackage{graphicx, psfrag}
\usepackage[textfont = small]{caption}
\usepackage{textcomp}
\usepackage{amsmath}
\usepackage{amsfonts}
\usepackage{gensymb}
\usepackage{stmaryrd}
\usepackage{dsfont}
 
\graphicspath{ {images/} }

\newcommand{\SMS}{Service Math\'ematiques et Statistiques}
\newcommand{\LSS}{Laboratoire des Signaux et Syst\`emes}
\newcommand{\CS}{CentraleSup\'elec}

\newcommand{\UPSUD}{Univ. Paris-Sud}
\newcommand{\FDS}{Fire Dynamics Simulator}
\newcommand{\NIST}{National Institute of Standards and Technology}

\newcommand{\twypg}{multi-fidelity non-stationary}
\newcommand{\TWYPG}{M-F$_{1}$}
\newcommand{\mfam}{multi-fidelity stationary}
\newcommand{\MFAM}{M-F$_{2}$}
\newcommand{\hf}{high-fidelity}
\newcommand{\HF}{H-F.}

\newcommand \vect[1]   {\boldsymbol{#1}}                     
\newcommand \matUn     {\mathds{1}}                          
\newcommand \ens[1]    {\mathbb{#1}}                         
\newcommand \esp[1]    {\mathbb{E}\left[#1\right]}           
\newcommand \var[1]    {\mathbb{V}\text{ar}\left[#1\right]}  
\newcommand \diag[1]   {\text{diag}\left\{#1\right\}}        
\newcommand \TC        {T^{c}}                               

\SetSymbolFont{stmry}{bold}{U}{stmry}{m}{n} 
 
\begin{document}

\begin{center}%
  {\Large%
    {\sc Gaussian process modeling for
      stochastic~multi-fidelity~simulators,
      with application to fire~safety}}%
  \bigskip
  
  R\'emi STROH $^{1, 2}$
  \& Julien BECT $^{2}$
  \& S\'everine DEMEYER $^{1}$
  \& Nicolas FISCHER $^{1}$
  \& Emmanuel VAZQUEZ $^{2}$
  \bigskip

  {\it $^{1}$ LNE, \SMS{};\\%
    29, avenue Roger Hennequin,%
    78190, Trappes; prenom.nom@lne.fr $^{2}$ \LSS{}\\%
    \CS{}, CNRS, \UPSUD{}, Universit\'e Paris Saclay\\%
    3, rue Joliot-Curie,%
    91190, Gif-sur-Yvette; prenom.nom@lss.supelec.fr }
\end{center}
\bigskip

{\bf R\'esum\'e.}  Pour évaluer les possibilités d'évacuation d'un
bâtiment lors d'un incendie, une méthode standard consiste à simuler
la propagation d'un incendie, au moyen de modèles de type différences
finies, et en prenant en compte le comportement aléatoire du feu, de
sorte que le résultat d'une simulation est non-déterministe.  La
finesse du maillage détermine la qualité du modèle numérique, ainsi
que son coût de calcul.  En fonction de la taille des mailles, une
seule simulation peut durer entre quelques minutes et quelques
semaines.  Dans cet article, nous cherchons à prédire le comportement
du simulateur à une maille fine, à partir de résultats moins coûteux,
à des mailles plus grossières.  Dans la littérature de la conception
et de l'analyse d'expériences numériques, on parle d'approche
multi-fidélité.  Notre contribution est d'étendre au cas de
simulateurs stochastiques du modèle bayésien multi-fidélité proposé
par Picheny et Ginsbourger (2013) et~\citet{tuo2014surrogate}.
\smallskip

{\bf Mots-cl\'es.} Exp\'eriences num\'eriques, Processus gaussien,
Multi-fid\'elit\'e, S\'ecurit\'e incendie \bigskip\bigskip

{\bf Abstract.}  To assess the possibility of evacuating a building in
case of a fire, a standard method consists in simulating the
propagation of fire, using finite difference methods and takes into
account the random behavior of the fire, so that the result of a
simulation is non-deterministic.  The mesh fineness tunes the quality
of the numerical model, and its computational cost.  Depending on the
mesh fineness, one simulation can last anywhere from a few minutes to
several weeks.  In this article, we focus on predicting the behavior
of the fire simulator at fine meshes, using cheaper results, at
coarser meshes.  In the literature of the design and analysis of
computer experiments, such a problem is referred to as multi-fidelity
prediction.  Our contribution is to extend to the case of stochastic
simulators the Bayesian multi-fidelity model proposed
by~\citet{picheny2013nonstationary} and~\citet{tuo2014surrogate}.

\smallskip


\section{Introduction}

\begin{table}[b]
  \begin{center}
    \begin{tabular}{|l|*{5}{|c}|}
      \hline
      Mesh size $t$ (cm)& 100 & 50 & 33.33 & 25 & 20\\
      \hline
      Duration of one simulation (h)& 1/12 & 1 & 6 & 20 & 54 \\
      \hline
    \end{tabular}
    \caption{Approximate duration of one run of \FDS{}, on the example
      presented in Section~\ref{sec:result}, as a function of the
      tuning parameter.}
    \label{tab:FDSduration}
  \end{center}
\end{table}

Fire Dynamics Simulator (FDS) is a numerical simulator developed by
the \NIST{}, that is used to simulate the propagation of fire in a
building, and assess its conformity to fire safety standards.  \FDS{}
is based on a finite difference method, that takes into account the
random behavior of fire propagation.  Consequently, the outputs of
\FDS{} are stochastic.  Using smaller mesh size increases the quality
of a simulation with respect to the physical reality, but also
increases the computational cost (see Table~\ref{tab:FDSduration}).
Thus, the mesh size controls a trade-off between speed and fidelity.
In other words, \FDS{} is a \emph{multi-fidelity} simulator.

Our work aims at estimating the behavior of \FDS{} at a very fine
mesh, with a limited computational budget, using the result of
simulations carried out using coarser mesh sizes.  To do this, we use
a Bayesian approach, where we construct a model of the output of
\FDS{} as a function of the mesh
size. Following~\citet{kennedy2000predicting} and others, our approach
is based on Gaussian process modeling.

Section~\ref{sec:model} shows how to extend Bayesian multi-fidelity
models proposed by~\citet{picheny2013nonstationary}
and~\citet{tuo2014surrogate} in the case of deterministic simulators,
to deal with the case of stochastic simulator.
Section~\ref{sec:result} presents numerical results to assess the
quality of our new model.

\section{Model for multi-fidelity}
\label{sec:model}

To formalize, consider $n$ input-output pairs of a stochastic
simulator with a tuning parameter, $\left( \left(\vect{x_i},
    t_i\right), Z_i\right)_{1\leq i \leq n} \in \left(\ens{X}\times
  \ens{T}\right)\times \ens{R}$, where $\ens{X}\subset\ens{R}^d$ and
$\ens{T}\subset\ens{R}^+$.  The outputs $Z_i$ are supposed to be
realizations of random variables, following distributions
$\text{P}_{\vect{x_i}, t_i}$.  The results are mutually independent.
In order to simplify, the distributions $\text{P}_{\vect{x_i}, t_i}$
are assumed to be Gaussian distributions, with means
$\xi\left(\vect{x_i}, t_i\right)$, and variances
$\lambda\left(\vect{x_i}, t_i\right)$~:
\begin{equation}
  Z_i\sim \mathcal{N}\left(\xi\left(\vect{x_i}, t_i\right),
    \lambda\left(\vect{x_i}, t_i\right)\right).
\end{equation}

To simplify, $\lambda$ is supposed to depend only on $t$~:
$\lambda\left(\vect{x}, t\right) = \lambda\left(t\right)$.  Besides,
we add a Gaussian prior distribution on the mean process, $\xi$. We
assumed that $\xi \sim GP(m, k)$. The prior distributions of $\xi$ and
$\lambda$ are independent.

A popular model for the process $\xi$ was developed
by~\citet{kennedy2000predicting}~: it is a recursive model, built in
the case of finite number of levels, which links two successive
Gaussian processes by a autoregressive relationship,
$AR\left(1\right)$.  However, this model is not well-suited to a
simulator with a continuous tuning parameter. Indeed, even if this
model can be extended for any number of levels~\citep{le2013multi},
the number of covariance parameters increases strongly with the number
of levels.  Also, this model does not actually use the value of the
tuning parameter, $t$.

For these reasons, another modeling was recently developed
by~\citet{picheny2013nonstationary} and~\citet{tuo2014surrogate}.
They supposed that an ideal simulator can be thought up by setting the
tuning parameter to an extreme value ($t = 0$).  Then, the process
$\xi$ can be written as
\begin{equation}
  \xi\left(\vect{x}, t\right) = \xi_0\left(\vect{x}\right)
  + \varepsilon\left(\vect{x}, t\right),
  \label{eq:proposedModel}
\end{equation}
where $\xi_0$ models this ideal simulator, and $\varepsilon$
represents a deterministic numerical error, independent of $\xi_0$,
which decreases when $t$ tends to $0$~: $\lim_{t \rightarrow 0}
\esp{\varepsilon\left(\vect{x}, t\right)^2} = 0$.

Following this decomposition, the covariance function of $\xi$, $k$,
is the sum of two covariance functions~: $k_{\xi_0}\left(\vect{x},
  \vect{x'}\right)$ $+$ $k_{\varepsilon}\left( \left(\vect{x},
    t\right), \left(\vect{x'}, t' \right)\right)$.  Two assumptions
are made~: first, the covariance function $k_{\varepsilon}$ is
separable, $k_{\varepsilon}\left( \left(\vect{x}, t\right),
  \left(\vect{x'}, t' \right)\right)$ $=$ $r\left(t,
  t'\right)k_{X\varepsilon}\left( \vect{x}, \vect{x'}\right)$, then,
the two spatial covariance functions, $k_{\xi_0}$ and
$k_{X\varepsilon}$ are stationary.  We choose anisotropic Mat\'ern
covariance functions for both.  The correlation $r$ is chosen as a
function of Brownian covariance function:~$r\left(t, t'\right) =
\min\left\{t, t'\right\}^L$, $L$ a positive real parameter.

The mean of $\xi$, $m$, is supposed constant, with an improper uniform
prior distribution on $\ens{R}$.

Finally, in order to improve the estimation of the observation
variances, particularly on costly levels, we add a prior distribution
on $\left(\lambda\left(t\right)\right)_{t \in \ens{T}}$.  This prior
distribution describes two ideas~: values of $\lambda\left(t\right)$
are not precisely-known a priori, so $\var{\lambda\left(t\right)}$ are
large; but the variances are alike, so
$\var{\lambda\left(t\right)\vline \lambda\left(t'\right)}$ are small.
Finally, a log-normal prior is chosen~:
\begin{equation}
  \left(\ln\left[\lambda\left(t\right)\right]\right)_{t\in \ens{T}} 
  \sim \mathcal{N}\left(\ln\left(\lambda_{prior}\right)\vect{1},
    \varsigma^2I_d + s^2\matUn\right),
  \label{eq:aPrioriVar}
\end{equation}
with $\lambda_{prior}$ equals to $1\%$ of the range of the output;
$\vect{1}$ is the vector of ones; $I_d$ the identity matrix; $\matUn$
the square matrix of ones; $s^2 = \ln\left(10\right)^2 \gg \varsigma^2
= \left(\ln\left(2\right)/3\right)^2$.

Finally, our \twypg{} model is built as follows~:
\begin{align}
  \begin{split}
    & \left(Z_i\right)_{1\leq i\leq n}\vline \xi, \left(\lambda
      \left(t\right) \right)_{t\in \ens{T}} \sim
    \mathcal{N}\left(\left(\xi\left(\vect{x_i},
          t_i\right)\right)_{1\leq i\leq n},
      \diag{\left(\lambda\left(t_i\right)\right)_{1\leq i\leq n}}\right);\\
    &\xi \sim GP\left(m, k\right);\\
    & m\left(\vect{x}, t\right) = m \sim \mathcal{U}_{\ens{R}};\\
    & k\left( \left(\vect{x}, t\right), \left(\vect{x'},
        t'\right)\right) = k_{\xi_0}\left(\vect{x} - \vect{x'}\right)
    + \min\left\{t, t'\right\}^L
    k_{X\varepsilon}\left(\vect{x} - \vect{x'}\right);\\
    &\left(\ln\left[\lambda\left(t\right)\right]\right)_{t\in \ens{T}}
    \sim \mathcal{N}\left(\ln\left(\lambda_{prior}\right)\vect{1},
      \varsigma^2I_d + s^2\matUn\right).\\
  \end{split}
  \label{eq:PGmod}
\end{align}
The parameter $m$ is integrated out analytically.  All other
parameters, $\left(\lambda \left(t\right) \right)_{t\in \ens{T}}$, $L$
and the hyper-parameters of $k_{\xi_0}$ and $k_{X\varepsilon}$, are
estimated by maximization of the joint posterior density (MAP
estimation).

\section{Numerical results}
\label{sec:result}

We consider a parallelepiped building, with two doors and two windows,
simulated with \FDS{}.  We study the maximal temperature in the
building, $\TC_t\left(\vect{x}\right)$, as function of $d = 8$ inputs
(external temperature, fire area\dots).

The model presented in Section~\ref{sec:model} is called \twypg{}
model and denoted by \TWYPG{}.  Our objective here is to compare it to
two other models.  The first model is similar to~\eqref{eq:PGmod}, but
uses a stationary anisotropic Mat\'ern covariance function on
$\ens{X}\times\ens{T}$.  This model, called \mfam{} model and denoted
by \MFAM{}, is a simplification of the \twypg{} model.  The second
model is built only from the most accurate level of the simulator.
This model, called \hf{} model and noted \HF{}, serves as a reference
value.

\begin{table}
  \centerline{
    \begin{tabular}{|l||c|c|}
      \hline
      Kind of design & Multi-fidelity design & High-fidelity design\\
      \hline
      \hline
      Property & Nested & Latin Hypercube\\
      \hline
      Observations &
      \begin{tabular}{l|*{5}{c}}
        $t$ (cm) & 100 & 50 & 33 & 25 & 20\\
        \hline
        $n_{points}$ & 270 & 90 & 30 & 10 &  0\\
      \end{tabular} &
      \begin{tabular}{l|*{5}{c}}
        $t$ (cm) & 100 & 50 & 33 & 25 & 20\\
        \hline
        $n_{points}$ &   0 &  0 &  0 &  0 & 100\\
      \end{tabular}\\
      \hline
      Speed & $\approx 11$ faster & 1 (reference)\\
      \hline
    \end{tabular}
  }
  \caption{Summary of designs used for comparison. 
    $n_{points}$~: number of points.}
  \label{tab:sumDesign}
\end{table}

Two datasets have been built for this numerical experiment (see
Table~\ref{tab:sumDesign}).  The first one is built with 400
simulations, at different mesh sizes.  It is used to build both
multi-fidelity models (\TWYPG{} and \MFAM).  The second one consists
of 100 simulations at $t = 20$~cm.  It is used to build the \hf{}
model \HF, and also for validation.

\begin{figure}
  \begin{tabular}{cc}
    \psfrag{Observations of TC}[tc][tc]{\hspace{0.0cm}\footnotesize
      ${\TC_{20cm}}_{obs}$}
    \psfrag{Predictions of TC}{\hspace{-1.4cm}\footnotesize
      Predictions of $\TC_{20cm}$ (\degree C)}
    \psfrag{PG model}[cb][cb]{\footnotesize{}\TWYPG{}}
    \psfrag{HM model}[cb][cb]{\footnotesize{}\MFAM{}}
    \psfrag{Kri model}[cb][cb]{\footnotesize{}\HF{}}
    
    \includegraphics[width = 0.45\textwidth]{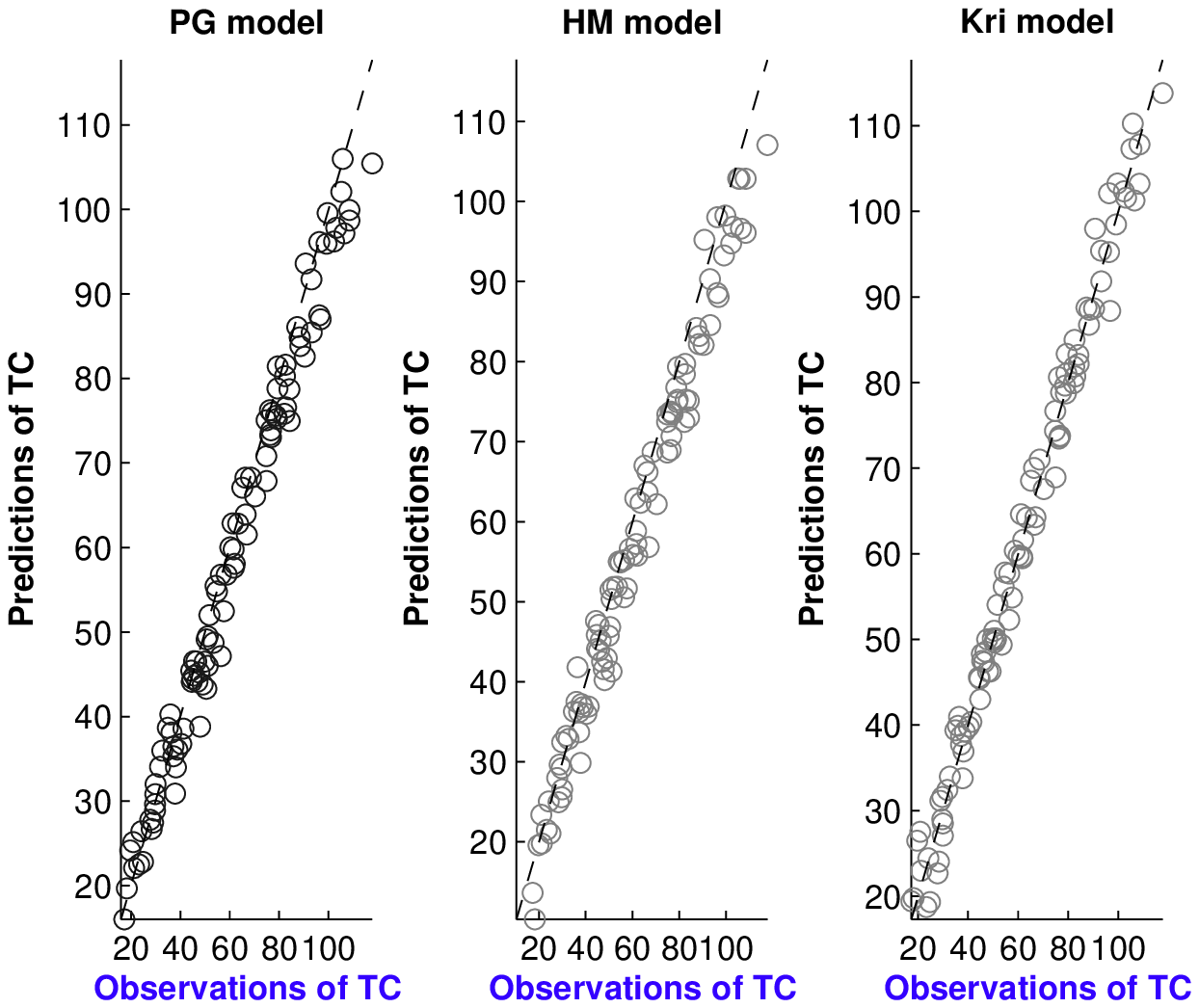}
    &
    \psfrag{TC normalized difference}[tc][tc]{\footnotesize{}
      $\Delta_{\TC_{20cm}}$}
    \psfrag{Estimation of the density function}{\hspace{-0.9cm}
      \footnotesize{}Probability density function}
    \psfrag{PG model}{\hspace{-0.1cm}\footnotesize{}\TWYPG{}}
    \psfrag{HM model}{\hspace{0.0cm}\footnotesize{}\MFAM{} }
    \psfrag{Kri model}{\hspace{0.1cm}\footnotesize{}\HF{}}
    
    \includegraphics[width = 0.45\textwidth]{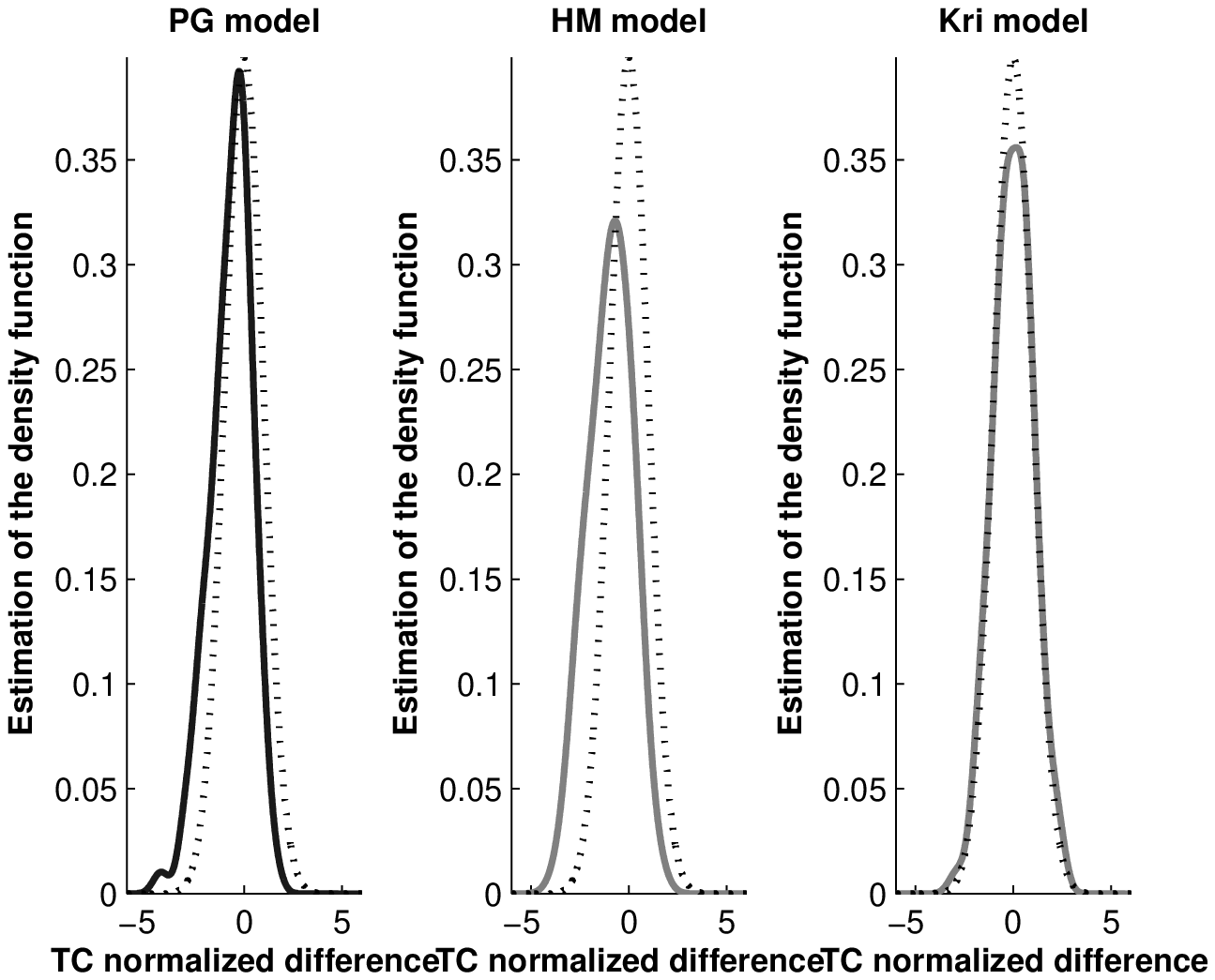} \\
  \end{tabular}
  \caption{Results on prediction of level 20 cm, by the three
    models. From left to right, \twypg{} model (\TWYPG); \mfam{} model
    (\MFAM{}); \hf{} model~(\HF).  Left~: predictions versus
    observations; dashed line~: first bisector ($y = x$).  Right~:
    Estimation of the probability density function of the normalized
    residuals $\Delta_{\TC_{20cm}}$; dashed line~: probability density
    function of normal distribution.}
  \label{fig:PredictResults}
\end{figure}

The results of prediction are presented on
Figure~\ref{fig:PredictResults}.  On the left, figures show a
comparison between predictions (posterior means) and observations.
For the \hf{} model, predictions are made by leave-one-out
cross-validation.  On the right, the densities of normalized residuals
are compared with the probability density function of the normal
distribution.  Overall, the two multi-fidelity models present a
goodness-of-fit similar to that of the \hf{} model, which is our
reference.  On closer inspection, it appears that both multi-fidelity
models---and most particularly the \mfam{} model---actually
underestimate $\TC$ for high values.  However, the standard deviations
of the densities of residuals are close to one, suggesting that
posterior variances are neither too large, nor too thin.

\begin{figure}
  \begin{center}
    \psfrag{P(TC > 60 C)}[tc][tc]{\footnotesize{}
      $p\left(\mathbb{P}_{\ens{X}}\left(\TC_{20cm} > 60 \degree
          C\right)\right)$}%
    \psfrag{Probability density function of the probability of
      failure}[cb][cb]{\footnotesize{} Posterior density}%
    \psfrag{Law of probability of failure of TC}{} \psfrag{nbSimu =
      1e3; nbPtsPerSimu = 5e3}{} \psfrag{PG}{\scriptsize{}\TWYPG{}}
    \psfrag{HM}{\scriptsize{}\MFAM{}} \psfrag{Kri}{\scriptsize{}\HF{}}

    \includegraphics[width = 0.5\textwidth]{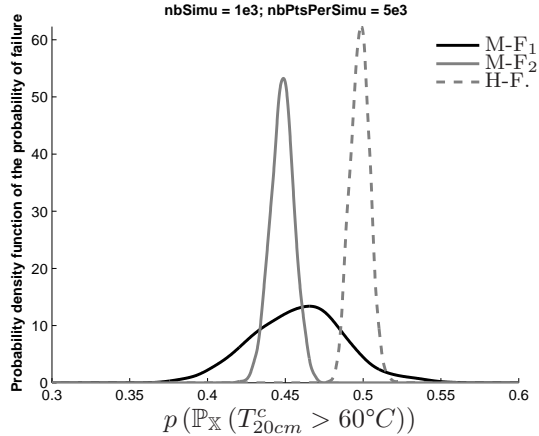}
    \caption{Estimations of the posterior function of
      $\mathbb{P}_{\ens{X}}\left(\TC_{20cm} > 60 \degree
        C\right)$. These densities are estimated with $n_{sim} = 1000$
      conditional random simulations on $n_{pts} = 5000$ inputs.}
    \label{fig:ProbaResults}
  \end{center}
\end{figure}

Finally, we consider the problem of estimating the probability
$\mathbb{P}_{\ens{X}}\left(\TC_{20 cm}\left(\vect{x}\right)> 60\degree
  C\right)$ that the output temperature exceeds a critical threshold
(here, $60\degree C$), where $\mathbb{P}_{\ens{X}}$ is a probability
distribution of the input space~$\ens{X}$.  Such a probability is
useful for fire engineers to assess the safety of the building.
Figure~\ref{fig:ProbaResults} presents different estimations of a
posterior probability density function of probability of exceeding the
threshold.  These densities are estimated with $n_{sim} = 1000$
conditional random simulations on $n_{pts} = 5000$ inputs.  The \hf{}
model yields the narrowest posterior distribution.  The \mfam{} model
also yields a small posterior uncertainty, but the support of the
density of the probability of exceeding the threshold does not agree
with that of the \hf{} model.  Our model has a larger posterior
variance, but is compatible with the reference model, and has been
obtained using less computational resources.

\section{Conclusion}

To conclude, we have proposed an extension of the model
of~\citet{picheny2013nonstationary} and~\citet{tuo2014surrogate} to
the case of stochastic simulators.  Our numerical results show that
the proposed model makes it possible to predict the behavior of a
stochastic multi-fidelity simulator at high fidelity, from simulations
at low fidelities.  We believe that this is a promising approach,
particularly in the domain of fire safety.  Future research will
concentrate on fully Bayesian estimation of the parameters of the
model, and sequential design of experiment in order to achieve a more
accurate estimation of the probability of exceeding the threshold,
using a limited computational budget for additional simulations.

\bibliographystyle{plainnat}
\bibliography{biblioArticleJDS}

\end{document}